\documentclass[letterpaper,11pt]{article}
\usepackage{jheppub,  tikz, physics, enumitem, braket, tikz-feynman, relsize, tensor, bbm}
\usepackage[mathscr]{eucal}
\usepackage[T1]{fontenc} 

\def\equationautorefname~#1\null{(#1)\null}

\tikzfeynmanset{compat=1.1.0}

\newcommand{\f}{\frac}
\def\({\left(}
\def\){\right)}

\newcommand{\lb}{\left[}
\newcommand{\rb}{\right]}

\def\inf{\infty}

\newcommand{\beq}{\begin{equation}}
\newcommand{\eeq}{\end{equation}}
\newcommand{\beqs}{\begin{equation*}}
\newcommand{\eeqs}{\end{equation*}}
\newcommand{\beqar}{\begin{eqnarray}}
\newcommand{\eeqar}{\end{eqnarray}}
\newcommand{\bal}{\begin{aligned}}
\newcommand{\eal}{\end{aligned}}
\newcommand{\bmult}{\begin{multline}}
\newcommand{\emult}{\end{multline}}

\newcommand{\bfy}{{\bf y}}

\newcommand{\bfp}{{\bf p}}
\newcommand{\bfq}{{\bf q}}

\DeclareMathOperator{\e}{e}

\def\d{\delta}
\def\De{\Delta}

\def\e{\epsilon}

\newcommand{\vp}{\varphi}

\def\t{\tau}
\def\k{\kappa}

\def\w{\omega}

\newcommand{\tih}{\tilde H}

\title{\boldmath Cosmology as a weak gravitational field and\\
the trans-Planckian problem}

\author{Ilia Komissarov,}
\author{Alberto Nicolis,}
\author{and John Staunton}
\affiliation{Center for Theoretical Physics and Department of Physics,\\ Columbia University, New York, NY 10027, USA}

\emailAdd{i.komissarov@columbia.edu}
\emailAdd{a.nicolis@columbia.edu}
\emailAdd{j.staunton@columbia.edu}

\abstract{At momenta much higher than the Hubble scale, the cosmological expansion can be thought of as a weak gravitational field. We consider QFT in a particularly convenient set of coordinates that makes this manifest, so that, for those high momenta, the effects of the cosmological expansion can be dealt with using the standard tools of perturbation theory in Minkwoski space. In this way, we re-derive standard results of QFT in a cosmological background, such as mode-stretching and gravitational particle production. We discuss the implications of our results for the trans-Planckian problem.}

\begin{document} 
\maketitle
\flushbottom

\section{Introduction}
\label{s:Introduction}

There has been a recent revival of the so-called trans-Planckian problem of inflationary cosmology \cite{MB}, mainly motivated by the associated ``trans-Planckian censorship'' conjecture \cite{BBLV}. 
The trans-Planckian problem can be simply stated as follows. Cosmological expansion stretches all wavelengths. Thus, with enough expansion --- in particular, with enough inflation --- it is possible that some of the modes that were shorter than the Planck scale sometime during inflation exited the horizon at some later time and can now be observed in the CMB or in large-scale structures. How can we trust our inflationary perturbation computations since they involve trans-Planckian physics?

Qualifying this as a ``problem'' is perhaps too pessimistic: the possibility that trans-Planckian physics has observable imprints in cosmological observables would constitute a unique opportunity to observe quantum gravity phenomena. However, regardless of how we name it, the claim that there are potentially observable effects at all has always been, understandably, a controversial one.

Several objections to the main logic behind the trans-Planckian problem have been raised. Our own viewpoint is that, as long as the Hubble rate during inflation is much smaller than the Planck mass, then, according to trans-Planckian degrees of freedom, the expansion of the universe is adiabatic. And so, if these start off in their ground state,
they will stay there up to exponentially small corrections. This should be guaranteed by the quantum mechanical adiabatic theorem (see e.g.~\cite{WeinbergQM}).

We say ``should be,'' because the way we usually do computations in cosmological perturbation theory does not make the adiabaticity of the expansion for short modes manifest at all. Especially for an inflationary cosmology, there are exponential redshift factors that apply to all modes, and so there is no obvious sense in which the expansion of the universe is a small effect. On the other hand, we know it should be so, at least for short length scales and short times. It is guaranteed by the equivalence principle and we implicitly use this fact all the time when we do computations for, say, the LHC without ever worrying about the present Hubble expansion. This viewpoint is shared by the authors of \cite{BdAQ}.

The situation is reminiscent of another trans-Planckian problem. In the standard computation of Hawking radiation from a black hole, one formally {\em has} to use modes that, close to the horizon, have wavelengths shorter than the Planck scale. How can we trust our predictions there? Polchinski famously solved the puzzle with his so-called nice-slice argument \cite{Polchinski}. The problem is in the coordinates that we are using to do the computation. In principle, it is possible to foliate the black-hole spacetime with ``nicer'' equal-time slices, whose geometry is non-singular and non-degenerate everywhere, and that reduce to the standard equal-time hypersurfaces asymptotically far away from the black hole. For instance, one could use coordinates that smoothly interpolate between the Kruskal ones close to the black hole and the Schwarzschild ones far away. In these coordinates, one can argue that nowhere will one have to invoke trans-Planckian modes to compute Hawking radiation. 

Polchinski's argument relies on a gedanken computation: to our knowledge, nobody has performed the Hawking radiation computation in ``nice'' coordinates, simply because such coordinates might be nice from a physical viewpoint, but are probably horrible to do explicit computations in as they completely hide one of the isometries of Schwarzschild spacetime, time translations. For instance, in the Kruskal-to-Schwarzschild coordinates proposed above, this isometry must smoothly interpolate between a 2D Lorentz boost \cite{HN} and the constant shift of the time variable. Clearly, at intermediate distances, it must take a complicated form.

We feel that the situation with the inflationary trans-Planckian problem is similar. If one were to use ``nicer'' coordinates in which the Hubble expansion is manifestly a small perturbation for short enough modes, the adiabatic theorem could then be applied, and one could argue that the short-mode vacuum stays empty, up to exponentially small effects. Our modest contribution to the debate is to propose such a set of coordinates and to use them to perform some simple explicit perturbative computations for a free scalar field coupled to non-dynamical gravity.

By ``perturbative'' we mean that, in these coordinates, we will be treating the Hubble expansion itself as a small gravitational field in Minkwoski space, and we will expand observables in it by applying the standard techniques of  QFT perturbation theory in  {\em flat spacetime}. For example, for a high-momentum particle, we will rederive the gravitational redshift of its energy and momentum by considering the scattering of the particle off the gravitational background field. Similarly, we will rederive cosmological particle production in the high-momentum limit by computing a vacuum-to-two particle transition probability. Finally, in the massless, conformally coupled case, we will compare the exact two-point function to its perturbative expansion in our coordinates, and find perfect agreement.

The same qualification about isometries that we mentioned above for the black-hole case applies here as well. In our coordinates, the spatial-translation isometry of FRW is not manifest, as it must be to allow the \emph{physical} momentum to redshift, and this  complicates computations substantially. Still, as we will see shortly, these coordinates have some nice technical properties that make the perturbative expansion simpler than one would expect for a more general coordinate system.

Thanks to these technical properties, regardless of their relevance for the trans-Planckian problem, our coordinates and the associated perturbative techniques might turn out to be useful for other applications as well. They are optimized to treat the cosmological expansion at sub-Hubble distance scales as a small perturbation and to carry out a perturbative expansion in it.

{\em Notation and conventions:} We will use natural units with $\hbar = c =1$ units and the mostly-plus metric signature throughout. We will keep the spatial dimensionality $n$ generic, but for some computations, we restrict to the simplest case, $n=1$.
Since we will be doing perturbation theory in Minkowski space, unless otherwise specified, when we talk about spatial distances and momenta we mean the {\em physical} ones, and not the comoving ones. So, in particular, we will be labeling single particle states $\ket{\vb{p}}$ by their physical momenta $\vb{p}$, and we will use the so-called relativistic normalization for them, 
\begin{equation}
\label{eq:relnorm}
    \braket{\vb{q}|\vb{p}} = \sqrt{2 \w_p} \sqrt{2 \w_q} \left(2\pi\right)^n \delta^n \left(\vb{p} - \vb{q}\right)\, ,
\end{equation}
where $\w_p$ and $\w_q$ are the energies of the particles with momenta $\bfp$ and $\bfq$.
\section{Physical coordinates}
\label{s:PhysicalCoordinates}

The set of coordinates we propose to use is directly related to what we usually call physical distances in cosmology:
\beq
\vb{y} = a(t)\vb{x},
\eeq 
where the $x^i$ are the standard FRW comoving coordinates. Moreover, we will use cosmic time $t$ as a time variable, which, as is well known, is the proper time of comoving observers. Because of all this, we will refer to this set of coordinates, $Y^\mu = (t, \vb{y})$, as {\em physical coordinates}.

The spatially flat FRW metric,
\beq
\label{eq:FRW}
d{s}^2 = -d{t}^2 + a^2(t) \, d{\vb{x}}^2 \; ,
\eeq
in physical coordinates simply becomes
\begin{align}
\label{eq:PhysCoords}
d{s}^2 & = -\left(1 - H^2(t)\, \bfy^2\right)d{t}^2 - 2 H(t) \, \bfy \cdot d \bfy  \, dt  + d \bfy \, ^2 \\
& = \big( \eta_{\mu\nu}+ h_{\mu\nu} \big) dY^\mu \, dY^\nu \; , 
\end{align}
where the perturbation field $h_{\mu\nu}=h_{\mu\nu}(Y)$ is given by
\beq
h_{00} = H^2(t) \, \bfy^2 \; , \qquad h_{0i} = -H(t) \, y^i \; , \qquad h_{ij} = 0 \;   \label{hmn}
\eeq
where, of course, $H(t)= \dot a/a$ is the Hubble rate.

A few comments are in order:
\begin{enumerate}
\item
For the de Sitter case, $H(t) = {\rm const.}$, this set of coordinates was introduced and used in \cite{Parikh}. In that case, they were dubbed {\em Painlev\'e-de Sitter} coordinates. Our coordinates can  thus be thought of as a generalization of those to more general cosmologies.
\item
Encouraged by the example of the de Sitter metric in static coordinates, one might be tempted to go one step further and redefine the time-variable as well, so as to put the metric in static-like form, with no off-diagonal $d{y}^i \, d{t}$ components. Although this is possible, we find it convenient not to do so, since it would completely spoil the simplicity of the perturbation $h_{\mu\nu} $ in \autoref{hmn}---in particular, its simple $\vb{y}$-dependence as well as the technical properties that we discuss at the end of this section
\item The perturbation in \autoref{hmn} is manifestly small for $H y \ll 1$, that is, at sub-Hubble distances from the origin, but at all times. We thus see that physical coordinates make the equivalence principle manifest and parametrize deviations from it in a relatively simple form, with a metric perturbation that stops at quadratic order in $ H y$.
\end{enumerate}

As for the technical advantage of these coordinates over others for perturbation theory: when we write down a QFT in curved spacetime, on top of the metric, we need its inverse, its determinant, the Christoffel symbols, etc. If then one wants to do perturbation theory in $h_{\mu\nu} = g_{\mu\nu} - \eta_{\mu\nu}$, most of these quantities will receive contributions of {\em any} order in $h_{\mu\nu}$ because the inverse of the metric does, and that enters  the definition of all of them (apart from the determinant of $g_{\mu\nu}$). This is not the case in physical coordinates as the inverse is simply
\beq
g^{\mu\nu} = \eta^{\mu\nu} + \delta g^{\mu\nu} \; ,
\eeq
with
\beq \label{eq:inverseh}
\delta g^{00} = 0, \qquad \delta g^{0i} = - H(t)\, y^i, \qquad \delta g^{ij} = - H^2(t) \, y^i y^j \; . 
\eeq
That is, the inverse of the metric also stops at quadratic order in $H y$. Moreover, the determinant of the metric is a constant,
\beq
g \equiv \det g_{\mu\nu} = -1 \; ,
\eeq
which incidentally shows that, despite the vanishing of $g_{00}$ at $|\bfy | = H^{-1}$, the metric is in fact non-degenerate and thus invertible everywhere. 

For what follows, it is also useful to display the Ricci scalar. This being a scalar, it is the same in any system of coordinates. In particular, it only depends on time, through $H(t)$ and its derivative,
\begin{equation}\label{eq:RicciS}
R = n\left(n + 1\right)H^2 (t) + 2n \, \dot{H} (t).
\end{equation}

\section{QFT in physical coordinates}\label{QFT physical}

To appreciate the technical virtues of physical coordinates for perturbation theory, consider for definiteness a free scalar field living in an FRW cosmological background:
\begin{equation}
    S = -\frac{1}{2}\int dt d^{n}{x} \sqrt{-g} \Big[ g^{\mu\nu} \partial_\mu \phi \partial_\nu \phi + m^2 \phi^2 + \xi R \, \phi^2 \Big] ,
\end{equation}
where we have introduced a generic coupling $\xi$ to the curvature scalar. 

In physical coordinates, the action simply reduces to
\beq
    S = S_{\rm free} + S_{\rm int} \; ,
\eeq    
where $S_{\rm free}$ is the action for the scalar in Minkowski space,
\beq
\label{eq:Sfree}
S_{\rm free} =
    -\frac{1}{2} \int dt d^n y  \, \Big[\eta^{\mu\nu} \partial_\mu \phi \partial_\nu \phi + m^2 \phi^2 \Big] \; ,
\eeq
and  $S_{\rm int}$ describes its interaction with the background gravitational field,
\begin{align}
\label{eq:Sint}
S_{\rm int}  = & -\frac{1}{2} \int dt d^n y \, \big(  \delta g^{\mu\nu} \partial_\mu \phi \partial_\nu \phi + \xi R \, \phi^2 \big)\\
= & \, \frac{1}{2} \int dt d^n y \, \Big[ 2 H(t) {y}^i \, \partial_i \phi \, \dot \phi  + H^2(t) (y^i \,  \partial_i \phi)^2 \\
& - \xi \big(n\left(n + 1\right)H^2 (t) + 2n \, \dot{H} (t) \big) \phi^2 \, \Big] \; ,
\end{align}
where we have made {\em no} approximations. The second-to-last line describes the minimal coupling of our scalar to gravity. The last line supplements that with a generic coupling to the Ricci scalar; for a minimally coupled scalar, it can be dropped.

As far as perturbative computations in momentum space go --- say $S$-matrix computations --- recall that for each vertex involving an external field (or source), the energy and momentum-conserving delta-functions have to be replaced with the Fourier transform of the external field itself, evaluated at the net energy and momentum flowing {\em out} of the vertex (that is, the energy and momentum provided to the vertex by the external field).

In our case, the external fields in the interaction vertices are simply powers or derivatives of $H(t)$ multiplying powers of $y^i$. Their Fourier transforms can then be expressed in terms of spatial-momentum delta-functions and their derivatives as well as the Fourier transform of $H(t)$, 
\begin{align}
H(t) \, y^i \quad & \to \quad   i \tilde H(\omega) \,  (2\pi)^n \partial_{k^i} \delta^n(\vb{k}) \\
H^2(t) \, y^i y^j \quad & \to \quad  -  (\tilde H * \tilde H)(\omega) \,  (2\pi)^n \partial_{k^i} \partial_{k^j}\delta^n(\vb{k}) \\
\dot H \quad & \to \quad  -i \omega  \tilde H(\omega) \,  (2\pi)^n \delta^n(\vb{k}) \\
H^2(t) \quad & \to \quad   (\tilde H * \tilde H)(\omega) \,  (2\pi)^n \delta^n(\vb{k}) \; ,
\end{align}
where with `$*$' we denote Fourier-space convolution, with measure $d \omega/(2\pi)$. The Feynman rules then are:
\begin{itemize}

\item
For {\bf minimal coupling}, we have two vertices: one of order $H$,
\begin{equation}
\label{eq:Vertex1}
\begin{aligned}
H(t) {y}^i \, & \partial_i \phi \, \dot \phi:  \\  &
    \begin{gathered}
\vspace{-0.7cm}
\feynmandiagram[layered layout, horizontal = a to c]{a -- [plain, momentum'= {$p$}] b [crossed dot, minimum size=3.5mm] -- [plain, momentum' = {$q$}] c};
\end{gathered} = \left(2\pi\right)^n\tilde{H}(\w_q - \w_p)\left(\w_p  q^i + \w_q  p^i \right)\partial_{i} \delta^n(\vb{q} - \vb{p}) \; ,
\end{aligned}
\end{equation}

and one of order $H^2$,
\begin{equation}
\label{eq:Vertex2}
\begin{aligned}
\frac12 H^2(t) & (y^i \,  \partial_i \phi)^2: \\ &
\begin{gathered}
\vspace{-0.7cm}
\feynmandiagram[layered layout, horizontal = a to c]
{a -- [plain, momentum'= {$p$}] b [empty dot, minimum size=3.5mm] -- [plain, momentum'={$q$}] c};
\end{gathered} = - i\left(2\pi\right)^n(\tilde{H} \ast \tilde{H})(\w_q - \w_p) \, p^i q^j \, \partial_i \partial_j \delta^n(\vb{q} - \vb{p}).
\end{aligned}
\end{equation}

\item
For {\bf non-minimal coupling}, we have two additional vertices: one of order $\dot H$,
\begin{equation}
\label{eq:Vertex3}
\begin{aligned}
- \xi n \, & \dot{H} (t)  \phi^2: \\ &
\begin{gathered}
\vspace{-0.7cm}
\feynmandiagram[layered layout, horizontal = a to c]
{a -- [plain, momentum'= {$p$}] b [dot, minimum size=3.5mm] -- [plain, momentum'={$q$}] c};
\end{gathered} = - 2 \xi n  \, (\w_q - \w_p) \, \tilde{H}(\w_q - \w_p) \, (2\pi)^n \delta^n(\vb{q} - \vb{p}) \; ,
\end{aligned}
\end{equation}

and one of order $H^2$,
\begin{equation}
\label{eq:Vertex4}
\begin{aligned}
-\frac12 \xi \, & n\left(n + 1\right)H^2 (t) \phi^2: \\ &
\begin{gathered}
\vspace{-0.7cm}
\feynmandiagram[layered layout, horizontal = a to c]
{a -- [plain, momentum'= {$p$}] b [square dot, minimum size=3.5mm] -- [plain, momentum'={$q$}] c};
\end{gathered} =
-i \xi n (n + 1) \, (\tilde{H} * \tilde{H})(\w_q - \w_p) \, (2\pi)^n  \delta^n(\vb{q} - \vb{p}) \; .
\end{aligned}
\end{equation}
\end{itemize}
We emphasize once again that  we will be doing standard flat-space perturbation theory. And so, in particular, the propagator to use is the standard Feynman one, as derived from the free action \autoref{eq:Sfree}:
\begin{equation}
\label{eq:propagator}
\begin{aligned}
\begin{gathered}
\vspace{-.8cm}
\feynmandiagram[layered layout, horizontal = a to b]
{a -- [plain, momentum'= {$p$}] b};
\end{gathered} \quad = \quad  \frac{i}{-p^2-m^2+i \epsilon} \; ,
\end{aligned}
\end{equation}
and the on-shell condition is the usual relativistic one
\beq
p^0 = \omega_p \equiv \sqrt{\vb{p}^2+m^2} \qquad \qquad \mbox{(on shell.)}
\eeq

A final technical remark is in order: when dealing with {\em derivatives} of delta-functions, one has to be particularly careful in ``using the delta,'' that is, in simplifying the form of whatever multiplicative function one has by making use of the fact that the delta-function only has support at vanishing argument. In order not to get confused, it is useful to always start from the distributional identity
\beq
f(x) \delta(x) = f(0) \delta(x) \; ,
\eeq
and take derivatives of both sides. By moving all terms that have $f$ and its derivatives only evaluated at $x=0$ to the r.h.s., one then gets useful distributional identities for the derivatives of the delta-function:
\begin{align}
\label{diracid}
f(x) \delta'(x) & = f(0) \delta'(x) - f'(0) \delta(x) \; , \\ 
 f(x) \delta''(x) & = f(0) \delta''(x) - 2 f'(0) \delta'(x) + f''(0) \delta(x)  \; ,
\end{align}
and so on. In the following, we will repeatedly use the $n$-dimensional generalizations of those identities:
\begin{align}
f(\vb{k}) \partial_{i} \delta^n(\vb{k}) & = f(\vb{0})  \partial_{i} \delta^n(\vb{k})  - \partial_{i} f(\vb{0})  \delta^n(\vb{k}) \; , \label{delta1} \\ 
f(\vb{k}) \partial_{i} \partial_j \delta^n(\vb{k}) & = f(\vb{0})  \partial_{i} \partial_j \delta^n(\vb{k}) 
- 2\partial_{(i} f(\vb{0}) \partial_{j)} \delta^n(\vb{k}) + \partial_{i} \partial_j f(\vb{0})  \delta^n(\vb{k})  \label{delta2} \; .
\end{align}
When $f$ involves the energy of an on-shell particle, the derivatives can act on that energy as well, in which case one has, as usual,
\beq
\frac{\partial \omega_p}{\partial p^i} = \frac{p^i}{\omega_p} \; .
\eeq

\section{Perturbative calculations to lowest order}
\label{s:PerturbativeCalcs}

We are now in a position to perform some simple perturbative computations and to check if they reproduce
what we know about QFT in a cosmological background. For the purposes of this section, we will consider expansions to first order in $H$ only. Since $H$ is dimensionful, naively this should correspond to expanding to the first order in $H/p$. In fact, as we will see, the systematics of the perturbative expansion is more subtle than that.

\begin{figure}[t]
\begin{center}
\includegraphics[width=14cm]{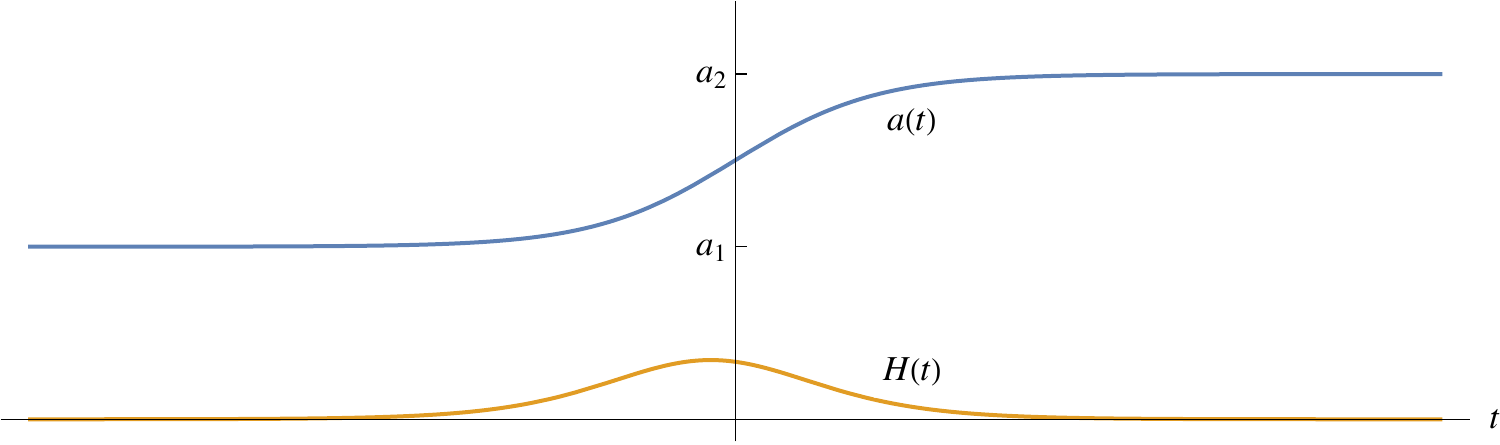}
\caption{\em The toy cosmology discussed in sect.~\ref{s:PerturbativeCalcs}. In the far past and far future, the scale factor is asymptotically constant and the associated Hubble rate drops to zero. \label{scale factor}}
\end{center}
\end{figure}

A technical but important caveat is that in order for us to have well-defined $S$-matrix elements, the interactions must go to zero in the infinite past and infinite future. Therefore, in our context, we must have $H(t) \to 0$  for $t \to \pm \infty$. This is equivalent to a spacetime where the scale factor approaches a constant ($a_1$) in the infinite past and a potentially different constant ($a_2$) in the infinite future, as depicted in \autoref{scale factor}. 
This is clearly not the cosmology of our universe, and it is also unrealistic for a more fundamental reason: it violates the null energy condition since it  necessarily involves a phase with positive $\dot H$. However, this spacetime is a useful toy model for studying quantum effects on a cosmological background and is often utilized in textbook treatments of the subject \citep{ParkerToms, BirrellDavies}. One can probably extend our techniques to more realistic cosmologies, ideally  with an inflationary phase in the past, using the standard QFT tricks of focusing on local-in-time probability rates rather than total probabilities, but we have not tried doing that yet. 

\subsection{Particle Production}
\label{s:ParticleProduction}

The first computation we handle is cosmological particle production. Within our framework, this simply corresponds to pair-production triggered by the time dependence of the external field --- in essence, the Schwinger effect for our scalar particles, with the gravitational field playing the role of the electromagnetic one in the case of QED.

We thus have to compute a vacuum-to-two particle transition probability. To first order in $H$, including the non-minimal coupling to curvature, there are two diagrams that contribute to the amplitude:
\begin{align}
    \mel{\vb{p}, \vb{q}}{S}{0} & \simeq  \begin{gathered}
\begin{tikzpicture}[rotate=45, transform shape]
\begin{feynman}
\diagram
{a [crossed dot, minimum size=3.5mm] -- [plain] b;
a -- [plain] c};
\end{feynman}
\end{tikzpicture}
\end{gathered}
\hspace{0.1cm}+\hspace{0.1cm}
\begin{gathered}
\begin{tikzpicture}[rotate=45, transform shape]
\begin{feynman}
\diagram
{a [dot, minimum size= 3.5mm] -- [plain] b;
a -- [plain] c};
\end{feynman}
\end{tikzpicture}
\end{gathered} \\
& = -\left(2\pi\right)^n\tilde{H}(\w_q + \w_p)\Big[(\w_p q^i + \w_q p^i)\partial_i \delta^n\left(\vb{q} + \vb{p}\right)+ 2n\xi\left(\w_q + \w_p\right)\delta^n\left(\vb{q}+ \vb{p}\right)\Big] \\ 
& = \tilde{H}(2 \w_p) \, \bigg[n \w_p (1-4\xi) - \frac{\vb{p}^2}{\w_p} \bigg] (2\pi)^n \delta^n\left(\vb{q}+ \vb{p}\right) \, \; ,
\end{align}
where $(p^i, \w_p)$ and $(q^i, \w_q)$ are the physical momenta and energies of the two outgoing particles, and in the last step we have ``used the deltas,'' as reviewed at the end of \autoref{QFT physical}.

The total transition probability, which is also the total average number of pairs produced, is formally
\begin{equation}
   {\cal N} = 
    \int \frac{d^n {y} d^n{y'}d^n{p}d^n{q}}{\left(2\pi\right)^{2n}} \frac{\abs{\mel{\vb{p}, \vb{q}}{S}{0}}^2}{\braket{\vb{p}, \vb{q}|\vb{p}, \vb{q}}\braket{0|0}} \; ,
\end{equation}
which has the usual infinite-volume divergence, ${\cal N} \sim {\cal V}$, but is free from the usual infinite-time divergence, thanks to our interactions shutting off at early and late times. Using standard regularization methods \cite{Weinberg1}, $(2\pi)^n \delta^n(\vb{0}) = \cal V$, we get  that the number of particles produced per unit phase-space volume at momentum $\vb{p}$ is\footnote{The number of particles with momentum $\vb{p}$ is the same as the number of particle pairs with momenta $\vb{p}$ and $-\vb{p}$.}
\beq \label{eq:MainParticle}
n_{\vb{p}} \equiv {d {\cal N}}\Big/
\frac{d^n y \, d^n p}{(2 \pi)^n} = \frac{1}{4}\bigg[\left(1 - 4\xi\right)n - \frac{\vb{p}^2}{\w_p^2}\bigg]^2 \left|\tilde{H}(2 \w_p)\right|^2 \; .
\eeq
%
%
This constitutes the main result of this subsection. We would like to stress how easy it was to derive, in full generality, for arbitrary $\xi$, $n$, and, especially, $a(t)$.

Note that for the  massless ($\w_p = |\vb{p}|$), conformally coupled ($\xi = (n-1)/4n$) case, the expected number of particles  vanishes. This matches the well-known fact that there is no cosmological particle production for conformally coupled massless fields \citep{ParkerToms, BirrellDavies}.

Note also that,  at this order,  particle production only depends on the cosmological history through  the Fourier transform of the Hubble parameter, $\tilde H(2 \w_p)$.\footnote{This result has some overlap with those of refs.~\cite{Gubser1, Gubser2}, and was also derived in perturbation theory in \cite{Tolley2005}. We thank A.~Tolley for making us aware of these earlier papers.}
 This happens to be dimensionless, and so can be used directly as a perturbation parameter, assuming that it is small. In particular, the high-momentum regime that we are interested in corresponds to having $H(t)$ not only much smaller than $p$ but also dominated by Fourier modes with frequencies much smaller than $p$. Then, $\tilde H(2 \w_p)$ will be extremely small, perhaps exponentially so.
As a bonus, from the computational viewpoint, computing or estimating the high-frequency limit of $\tilde H(\omega)$ is much less demanding than computing cosmological particle production in the usual way, which involves finding solutions to the equation of motion --- something that very rarely can be done analytically.

We can compare our result with what one finds by applying the standard techniques of Bogolyubov transformations \cite{ParkerToms, BirrellDavies}. As we just mentioned, the cases where such a procedure can be carried out analytically are few and far between. We analyze some cases in \autoref{app:TextbookResults} and only quote the results here. In what follows, let $d{\tau} = d{t}/a^n$, once again where $n$ is the number of spatial dimensions. For now, let's restrict ourselves to the minimally coupled ($\xi=0$), massless ($m=0$) case.

As a first example, consider the scale factor
\begin{equation}
\label{eq:ParticleEx1}
    a^{2\left(n - 1\right)}\left(\tau\right) = \frac{a_2^{2\left(n - 1\right)} + a_1^{2\left(n - 1\right)}}{2} + \frac{a_2^{2\left(n - 1\right)} - a_1^{2\left(n - 1\right)}}{2}\tanh\left(\frac{\tau}{2s}\right) \; ,
\end{equation}
where $a_1$ and $a_2 = a_1 + \Delta a$ are the two asymptotic values of $a$, and $s$ is a large `slowness' parameter.
It is possible to solve the equation of motion in the massless, minimally coupled case, find the corresponding Bogolyubov coefficients, and expand them to first order in $\Delta a/a_1$ as shown in \autoref{app:TextbookResults}. Writing the result in \autoref{eq:exactex1} in physical coordinates, we find the occupation number to be
\begin{equation}
\label{eq:ParticleEx1Final}
  n_{\vb{p}}  = \abs{\beta_p}^2 \simeq \frac14 \bigg[ (n - 1) \frac{2\pi s a_1^n \, p}{\sinh\big(2\pi s a_1^n \, p\big)}\frac{\Delta a}{a_1} \bigg]^2.
\end{equation}
On the other hand, using our method, we can first expand the Hubble parameter, $H(t)$, to first order in $\Delta a/a_1$. This yields
\begin{equation} \label{Hubble1}
    H(t) \simeq \frac{1}{4sa_1^n}\frac1{\cosh^2\left(\frac{t}{2sa_1^n}\right)}\frac{\Delta a}{a_1}   \qquad
    \Rightarrow \qquad 
    \tilde H(\omega) \simeq \frac{\pi s a_1^n \, \omega}{\sinh(\pi s  a_1^n \, \omega)} \frac{\Delta a}{a_1} \; .
\end{equation}
Plugging in the relevant parameters, $m = 0$ and $\xi = 0$, into \autoref{eq:MainParticle}, we find perfect agreement with \autoref{eq:ParticleEx1Final}.

As a second example, consider the scale factor,
\begin{equation}
\label{eq:ParticleEx2}
    a^{2\left(n - 1\right)}\left(\tau\right) = a_1^{2\left(n - 1\right)} + \frac{b}{4 \cosh^2\left(\frac{\tau}{2s}\right)} \; ,
\end{equation}
where $s$ is once again a large slowness parameter, and $b$ measures the overall size of the `bump' in $a(\tau)$, as it returns to its original value $a_1$ in the infinite future. This time the occupation number can be expanded to the lowest order in $b$ to obtain \autoref{eq:exactex2} and then written in terms of the physical momentum,
\begin{equation}
\label{eq:ParticleEx2Final}
    n_{\vb{p}} \simeq \frac14  \bigg[\frac{2 \pi s^2 a_1^2 \,  p^2}{ \sinh\left(2\pi s a_1^n p \right)} b \bigg]^2 \; .
\end{equation}
On the other hand, the Hubble rate to first order in $b$ is given by
\begin{equation}
    H(t) \simeq \frac{b}{8\left(n - 1\right)s a_1^{3n - 2}}\frac{\tanh\left(\frac{t}{2sa_1^n}\right)}{\cosh^2\left(\frac{t}{2s a_1^n}\right)}  \qquad \Rightarrow \qquad \tilde H(\omega) \simeq - \frac{i}{2(n-1)} \frac{\pi s^2 a_1^2 \,  \omega^2}{ \sinh\left(\pi s a_1^n \omega \right)} b\; .
\end{equation}

Plugging this into \autoref{eq:MainParticle} with $\xi= 0$, $m=0$ yields perfect agreement with \autoref{eq:ParticleEx2Final}\footnote{Notice that for a scalar field in $1+1$ dimensions, minimal coupling and conformal coupling happen to be the same. So, for $n=1$, $m=0$, and $\xi=0$, there should be no particle production, as correctly predicted by our formula \autoref{eq:MainParticle}. The fact that in this particular example we seem to be violating this property --- see eq.~\autoref{eq:ParticleEx2Final} --- is due to the Hubble rate's diverging for $n\to1$.}.

As a third and final example, consider a minimally coupled scalar field now with nonzero mass $m$, but only in $n = 1$ spatial dimensions, as the equation of motion for a massive particle does not have a known analytic solution for general $n$. Consider the scale factor,
\begin{equation}
    a^2\left(\tau\right) = \frac{a_2^2 + a_1^2}{2} + \frac{a_2^2 - a_1^2}{2} \tanh\left(\frac{\tau}{2s}\right).
\end{equation}
The Bogolyubov coefficients have been found in ref. \cite{BirrellDavies}. To lowest order in $\Delta a/a_1$, and in terms of the physical momentum, one gets
\begin{equation}
    n_{\vb{p}} =  \left[\frac{\pi s a_1 \, m^2}{\omega_p \sinh\left(2\pi s a_1 \, \omega_p \right)} \frac{\Delta a}{a_1} \right]^2 \;, \qquad \omega_p=\sqrt{\vb{p}^2+m^2} \; .
\end{equation}
For our method, the Hubble parameter $H(t)$ is still given by \autoref{Hubble1}, except the Fourier transform is now evaluated at $\omega = 2\omega_p$ instead of $2p$ as it was in the massless case. Again, we find exact agreement with the textbook result above.

\subsection{Gravitational Redshift}
\label{s:GravitationalRedshift}

The second calculation we set out to do in perturbation theory, again for a cosmology like that in \autoref{scale factor}, is that of gravitational redshift. Classically, this corresponds to the stretching of modes. Quantum mechanically, for a single particle, the problem can be phrased in terms of a $1 \rightarrow 1$ transition amplitude: what is the probability that a particle that started off with physical momentum $\vb{p}$ at $t=-\infty$ ends up with physical momentum $\vb{q}$ at $t=+ \infty$? We know the answer: the probability should be one if the final momentum is the correctly redshifted one, $\vb{q} = \vb{p} \, a_1/a_2$, and zero otherwise. However, as a function of $\vb{q}$, such a probability distribution is not particularly meaningful---for continuous variables, Kronecker-deltas should get replaced by Dirac-deltas. 

To get more meaningful results, one could use wave packets, or, more simply, restrict to studying the transition amplitude rather than the transition probability. Indeed, given that our states have delta-function normalization, up to finite normalization factors and a phase we expect
\beq
\mel{\vb{q}}{S}{\vb{p}} \propto (2\pi)^n \delta^n(a_2 \vb{q} - a_1 \vb{p}) \; .
\eeq
The full result, which we derive in \autoref{app:GravitationalRedshift} using QFT in curved spacetime is in fact
\beq \label{full redshift}
\mel{\vb{q}}{S}{\vb{p}} = e^{-i \varphi} \sqrt{(2 \omega_p)(2 \omega_q)}\, (a_1a_2)^{n/2} (2\pi)^n \delta^n(a_2 \vb{q} - a_1 \vb{p}) \; ,
\eeq
where the phase $\varphi$ can be computed in the adiabatic limit we are interested in as
\beq \label{phase}
\varphi \simeq  \int_{-\infty}^\infty dt \bigg[\omega(t) - \frac{\omega_p+ \omega_q}{2}\bigg] \qquad\qquad (q,p \gg H) \; ,
\eeq
and $\omega(t)$ is the instantaneous WKB energy of our particle, taking into account the gradual redshift of its momentum:\footnote{For simplicity, we are considering the minimally-coupled case, $\xi =0$. For non-minimal couplings, the $m^2$ term in $\omega(t)$ gets replaced by $m^2+ \xi \big(n\left(n + 1\right)H^2 (t) + 2n \, \dot{H} (t) \big)$. This changes our first-order formula \autoref{first order phase} for the phase  by a term that integrates to zero, as explained in the following footnote.}
\beq \label{omega(t)}
\omega(t) \equiv \sqrt{\frac{a_1^2}{a^2(t)} \vb{p^2}+m^2} = \sqrt{\frac{a_2^2}{a^2(t)} \vb{q^2}+m^2} \; .
\eeq

Although we postpone deriving it until the appendix, notice that the full result \autoref{full redshift} makes a lot of sense: it includes the expected redshift of momentum, it is invariant under an overall rescaling of $a(t)$, it includes the energy prefactors associated with our relativistic normalization of states, and, as far as the phase goes, it takes into account the difference between the adiabatic time evolution $e^{-i \int\!  \omega(t) \, dt}$ and the unperturbed ones, $e^{-i \omega_p t}$ and $e^{-i \omega_q t}$. The question for us is whether we can recover it with our perturbation theory in flat space.

To begin with, notice that the phase $\varphi$ depends on the whole history $a(t)$, but everything else only depends on the two endpoints $a_1$ and $a_2$, or, given the invariance under rescalings of $a(t)$, only on the total expansion factor $a_2/a_1$.  This is
\beq
\frac{a_2}{a_1} = e^{\int_{-\infty}^\infty dt \, H(t)} = e^{\tilde H(0)} \; .
\eeq
And so, expanding the full result \autoref{full redshift} in the total number of $e$-folds is the same as expanding in $\tilde H(0)$!
Clearly, there is a connection with our perturbative expansion, which involves, in each vertex, the Hubble rate's Fourier transform $\tilde H(\omega)$.
The expansion of the $e^{-i\varphi}$ phase factor in powers of $H$ will certainly be more complicated, especially at high orders. So, let's see how things work out at first order.

We want to compute the transition amplitude above in perturbation theory. To first order in $H$ we have the diagrams in \autoref{eq:Vertex1} and \autoref{eq:Vertex3}. The latter however does not contribute, since the delta function and the on-shell condition enforce $\w_q-\w_p=0$. We thus get,
\begin{align}
    \mel{\vb{q}}{S}{\vb{p}} = \: &  2 \w_p \, (2\pi)^n  \delta^n\left(\vb{q} - \vb{p}\right) \\
    &  +  \tilde{H}(\w_q - \w_p) \, (\w_p q^i + \w_q p^i ) \, (2\pi)^n \partial_i \delta^n\left(\vb{q} - \vb{p}\right) \; ,
\end{align}
where the first line is the zeroth order result --- the famous `1' in the $S$-matrix --- and the second line is the first-order contribution. We now use the identity \autoref{delta1}, interpreting the derivative as being with respect to $\vb{q}$ and concentrating only on the first order:
\begin{align}
\label{firstordershift1}
    \mel{\vb{q}}{S}{\vb{p}}^{(1)} = &\: \left[2 \w_p p^i \left(2\pi\right)^n \partial_i \delta^n\left(\vb{q} - \vb{p}\right) - \left(n \w_p + \frac{\vb{p}^2}{\w_p}\right) \left(2\pi\right)^n \delta^n\left(\vb{q} - \vb{p}\right)\right]\tilde{H}\left(0\right) \\
    &- 2\tilde{H}'\left(0\right)\vb{p}^2 \left(2\pi\right)^n \delta^n\left(\vb{q} - \vb{p}\right).
    \label{firstordershift2}
\end{align}
We can classify the various terms into those that involve $\tilde{H}(0)$, as in the first line, and those that involve $\tilde{H}'(0)$, as in the second line. We can then check if these match \autoref{full redshift} to first order in $\tilde{H}$.

Since $H(t)$ is real, $\tilde H(- \omega) = \tilde H^*(\omega)$, and so  $\tilde H(0)$ is real and $\tilde H'(0)$ is imaginary. As a consequence, at this order $\tilde H'(0)$ can only contribute to a phase factor. Ignoring phase factors for the moment, let's rewrite the r.h.s.~of \autoref{full redshift} as
\begin{align}
& \sqrt{(2 \omega_p)(2 \omega_q)}\,  (2\pi)^n \delta^n\bigg(\sqrt{\frac{a_2}{a_1}} \vb{q} - \sqrt{\frac{a_1}{a_2}} \vb{p} \bigg) \\
& = \sqrt{(2 \omega_p)(2 \omega_{p\,\cdot\, e^{-N}})} (2\pi)^n \delta^n\bigg(e^{N/2} \vb{q} - e^{-N/2} \vb{p} \bigg) \; ,
\end{align}
where $N $ is the total number of $e$-folds. Expanding in $N = \tilde H(0) $ we get the first order terms
\beq
\tilde H(0) \bigg[ -  \frac{\vb{p}^2}{\w_p}  (2\pi)^n  \delta^n\left(\vb{q} - \vb{p}\right) + \w_p (q^i+p^i)  \, (2\pi)^n \partial_i \delta^n\left(\vb{q} - \vb{p}\right) \bigg] \; ,
\eeq 
which, upon using \autoref{delta1} for the $\partial \delta$ term, exactly match \autoref{firstordershift1}.

As for the phase factor, notice that we can rewrite $\varphi$ in \autoref{phase} as
\begin{align}
\varphi & =  \frac12 \int_{-\infty}^\infty dt \big[(\omega(t) - \omega_p) + (\omega(t)- \omega_q)\big]\\
& = \frac12 \int_{-\infty}^\infty dt \bigg[ \int_{-\infty}^t dt' \, \dot \omega(t') 
 +\int_{\infty}^t dt'\, \dot \omega(t') 
\bigg] \\
& = \frac12 \int dt dt' \, {\rm sign}(t-t') \, \dot \omega(t') 
\; ,
\end{align}
where both time integrals now run from $-\infty$ to $+\infty$. Regulating the one in $t$ by restricting it to some large interval $[-T/2, T/2]$, and performing it first, leaves us with
\beq \label{phase t omegadot}
\varphi  = - \int dt' \, t' \, \dot \omega(t')  \; .
\eeq 

%
To first order in $H$, using $a(t) \propto \exp(\int dt\, H(t) )$, this is\footnote{For non-minimal couplings, to first order in $H$ there is an additional term proportional to
\beq 
\int dt' \, t' \, \ddot H(t') = - \int dt' \, \dot H(t') = 0 \; , 
\eeq
where we used that $H(t)$ and its derivatives go to zero at infinite times.
}
\beq \label{first order phase}
\varphi^{(1)} \simeq  \frac{\vb{p^2}}{\w_p} \int dt' \, t' \, H(t') = - i \, \frac{\vb{p^2}}{\w_p} \tilde H'(0)
\; ,
\eeq
which, plugged into \autoref{full redshift}, yields exactly the first-order phase \autoref{firstordershift2}.

\section{Higher orders}
\label{s:HigherOrders}

We now consider pushing our perturbative expansion to higher orders. This is particularly relevant for the gravitational redshift case since the expansion parameter there is $\tilde H(0)$, which, as we showed, is the total number of $e$-folds. Clearly, even for short-wavelength modes that never leave the horizon, there can be large secular effects that build up over many $e$-folds --- most notably, the total redshift factor $e^N$. For large $N$, or even just $N \sim 1$, one wants to have a resummation of such effects.

There are three main obstructions to analyzing the perturbative expansion order by order. The first is that our interactions are not all of the same order --- some are of order $H$ while others are of order $H^2$ --- thus making keeping track of which combinations of interactions contribute at which order messy. This issue can be solved by introducing an auxiliary field, which replaces the second-order vertices with a propagator connecting two first-order ones. This doubles the number of fields but makes order-counting straightforward. We explain this procedure in \autoref{app:auxiliary}.
The second obstruction comes from the fact that our interactions do not conserve momentum or energy, and so with each new vertex, there is another momentum and energy integral. Luckily there are also many delta functions that can get rid of the momentum integrals, but the energy integrals remain. Moreover, the momentum delta-functions are derived, which brings us to the third obstruction: using the  distributional identities for derivatives of delta functions, eqs.~\autoref{delta1}, \autoref{delta2} quickly becomes messy if there are several derivatives of deltas multiplying several functions of momenta. In fact, we saw already at first order that checking that our perturbative results reproduced the correct ones required some work.

Ideally, one would like to have a simple bookkeeping procedure, but we have not been able to elaborate a general one yet. Things are, in fact, simpler when we look at the {\em resummation} of the perturbative series, as we do in the next section, but for now, let us try to simplify things as much as possible and look at the perturbative series in more detail. To this end, we will restrict ourselves to studying a massless, minimally coupled scalar in $1+1$ dimensions:
\beq \label{simple}
m= 0 \;, \qquad \xi = 0 \; , \qquad n=1 \; .
\eeq
Additionally, given the fact that minimal coupling and conformal coupling coincide in $1+1$ dimensions, this is arguably the simplest case to consider.

Restricting the full one-to-one transition amplitude \autoref{full redshift} to this case we find
\begin{align}
\mel{q}{S}{p} & = e^{-i \varphi} \, 2 \sqrt{pq} \, \
\sqrt{a_1a_2} \, (2\pi)
 \delta(a_2 q - a_1 p), \\
 & = 2 e^{-i \varphi} \, (2\pi)
 \delta\bigg(\sqrt{\frac{a_2 q}{a_1 p}} -  \sqrt{\frac{a_1 p}{a_2 q}} \, \bigg),
\end{align}
where for simplicity we are taking the spatial momenta $p $ and $q$ to be positive, so that $\omega_p = p$ and $\omega_q =q $. At this point, it is useful to parametrize the momenta and the asymptotic scale factors in an exponential way:
\beq \label{parametrization}
a_1 = A \, e^{-N/2} \; , \qquad a_2  = A \, e^{N/2} \; , \qquad  p = Q \, e^{\kappa/2} \; ,\qquad q = Q \, e^{-\kappa/2}  \; ,
\eeq
where $A$ is a common scale factor, $N = \tilde H(0)$ the total number of $e$-folds, $Q$ a common momentum scale, and $\kappa$ a (logarithmic) measure of momentum redshift.
In these variables, the amplitude simply becomes
\begin{align}
\mel{{q}}{S}{{p}} & = 2 e^{-i \varphi} \, (2\pi) \delta \bigg( 2 \sinh\frac{\kappa - N}{2}\bigg) \\
& = 2 e^{-i \varphi} \, (2\pi) \delta (\kappa - N ) \; . \label{1+1massless}
\end{align}

Moreover, the phase $\varphi$ is given by \autoref{phase t omegadot} with
\beq
\omega(t) = \frac{a_1}{a(t)} p = \frac{a_2}{a(t)} q = \frac{\sqrt{a_1 a_2 \, p q}}{a(t)} = Q \frac{A}{a(t)} \; .
\eeq
Similarly, the scale factor $a(t)$ can be written as
\begin{align}
a(t) & = a_1 \exp{ \int_{-\infty}^t dt' \, H(t') } = a_2 \exp{-\int_t^\infty dt' \, H(t')}\\
&  = A \exp{ \frac12 \int dt' \, {\rm sign}(t - t') H(t') } \; ,
\end{align}
where the last integral runs from $-\infty$ to $+\infty$.

So, putting everything together, we have
\beq \label{final amplitude 1+1}
\mel{{q}}{S}{{p}} = 2 \exp{ - i Q \int dt \, t H(t) e^{-\frac12 \int dt' \, {\rm sign}(t - t') H(t')}} \, (2\pi) \delta (\kappa - \tilde H(0) ) \; .
\eeq

Notice that the expansion of the delta function in powers of $\tilde H(0)$ still gives us derivatives of deltas,
\beq \label{delta n}
\delta (\kappa - \tilde H(0) ) = \sum_n \frac{(-1)^n}{n!}  \tilde H^n(0) \, \delta^{(n)} (\kappa)
\eeq
but now the prefactors are independent of the argument of the deltas, $\kappa$, and so we don't need to use the identities \autoref{delta1}, \autoref{delta2}, or their higher-order analogs.  As a result, the perturbative expansion should be much cleaner if phrased in these momentum variables ($Q$ and $\kappa$) rather than the original ones ($p$ and $q$).

Additionally notice that, when expanding the above amplitude in powers of $H$, we have two fundamental `building blocks':
\beq
\tilde H(0) = \int \frac{d \omega}{ (2\pi)} \, (2\pi) \delta(\omega) \, \tilde H(\omega) 
\eeq
and
\beq
\int dt'\, {\rm sign}(t - t') H(t') = \int \frac{d \omega}{ (2\pi)}  \,  2i \, P \( \frac {1} \omega \) \,  \tilde H(\omega)  e^{- i \omega t}
\eeq
where `$P$' denotes the principal value. Recalling the distributional identity,
\beq
\frac{1}{\omega + i \epsilon} = P \(\frac 1 \omega \) - i \pi \delta(\omega) \; ,
\eeq
one is led to conjecture that the expansion of \autoref{final amplitude 1+1} in powers of $H$ will involve multiple frequency-integrals of products of $\tilde H(\omega)$ and $i/(\omega+ i \epsilon)$ factors.

Alas, things are not as simple if we start from the free action and imagine doing perturbation theory as described in \autoref{QFT physical}. We still lack a simple characterization of the generic $j$-th order term  in terms of the improved $(Q, \kappa)$ variables described above. The generic diagram will be a chain of vertices of the form \autoref{eq:Vertex1} or \autoref{eq:Vertex2} connected by Feynman propagators \autoref{eq:propagator}, integrated over all intermediate momenta and energies. It appears that, after some manipulations, the $j$-th order contribution to our transition amplitude can be cast into the form
\beq
\begin{aligned}
  2 p q   & \int  \bigg( \prod\limits_{a=1}^{j-1} \frac{ d\omega_a d k_a}{(2 \pi)^2} \frac{ i k_a}{\omega_a - k_a + i \epsilon \, {\rm sign}(k_a)}\bigg)   \\
& \times  \tilde H(q - \omega_{j-1})  \tilde H(\omega_{j-1} - \omega_{j-2}) \dots \tilde H(\omega_1 - p)   \\
& \times (2 \pi)^j \d'(q-k_{j-1}) \d'(k_{j-1} - k_{j-2}) \dots \d'(k_{1} - p )  \; ,  
\end{aligned}
\label{general n-th order}
\eeq
which, however, we have verified only up to fourth order ($j=4$).

Even accepting the general form \autoref{general n-th order}, the problem with performing the $k_a$ integrals using the delta functions now has to do with the fact that the $\delta'$'s are all `linked' together, in the sense that each $k_a$ variable appears in two of them, thus making the integrals generate more and more derivatives of delta functions. This is precisely what we want---see eq.~\autoref{delta n}---but the bookkeeping is complicated. One can check that going to relative momentum variables, $\Delta k_a \equiv k_a-k_{a-1}$, does not simplify things much.

In \autoref{App1to1PertCom} we verify that the second-order term matches what one gets from expanding the full answer \autoref{final amplitude 1+1}, but an all-order check still eludes us. It is actually simpler to deal with resummations, as we now explain.

%

\section{Resumming the two-point function}
\label{2ptMatch}
Let us consider the $T$-ordered two-point function:
\beq \label{2pt}
D(y,y'; t,t') \equiv \braket{T \phi(y,t) \phi(y',t')} \; ,
\eeq
where for simplicity we are still considering the simple case of $1+1$ dimensions. 
Notice that, since our interactions break translational invariance in space and time, the two-point function depends separately on the two spacetime points, and not just on the difference in their coordinates. 

For arbitrary distances in space and time, we do not in general know what the two-point function looks like, because computing it would be equivalent to exactly solving the field equations, which we cannot do for a generic cosmology. However, for short {\em physical} spatial distances, but for any time separation, we could use the adiabatic approximation described in \autoref{app:TextbookResults}, and expand the field operator in the orthonormal modes
\beq
u^{\pm}_{{k}}(y,t) = \frac{1}{\sqrt{2\omega(t)a(t)}} \, e^{\mp i \int^t dt' \, \omega(t')} \, e^{ \pm i \frac{{k} \cdot {y}}{a(t)}} \; ,
\eeq 
where $k$ stands for {\em comoving} momentum.  Plugging such an expansion into \autoref{2pt}, one finds an approximate integral representation for $D$ valid at distances shorter than the Hubble radius. 

Such an approximation becomes {\em exact} in the massless conformally coupled case, eq.~\autoref{simple}. In fact, just using that the FRW metric is conformally flat and that in $(1+1)$-dimensions a scalar has vanishing conformal weight zero, it is immediate to show that
\beq
D(y,y'; t,t') \Big|_{m=0, \, \xi = 0} = \int \frac{d \omega d k }{(2 \pi)^2} \frac{i}{\omega^2 - k^2 + i \epsilon} e^{- i \omega [\t(t') - \t(t) ]} e^{ i k \left[\f{y'}{a(t')} - \f{y}{a(t)} \right]} \, ,
\label{PropFull}
\eeq
which is nothing but the Minkowski two-point function in terms of the comoving spatial coordinates $x(y,t) = y/a(t)$  and of conformal time,
\beq
\tau(t) \equiv \int^t \frac{dt'}{a(t')} \; . 
\eeq
One can check that \autoref{PropFull} obeys our $m=0$, $\xi =0$ equation of motion at non-coincident points, as it should.

The above two-point function was derived using standard properties of conformal field theories. 
The goal is to now demonstrate that if we were to write down the fully dressed propagator by resumming the series found in perturbation theory, we would rederive the exact two-point function, \autoref{PropFull}. In fact, for what follows we will not need the explicit form of the two-point function, but just that it obeys the correct equation of motion, and this we can write down exactly in complete generality. 

To make manipulations more straightforward, let's introduce a simple notation. Following standard functional methods \cite{Weinberg1}, we rewrite the free and interaction actions \autoref{eq:Sfree}, \autoref{eq:Sint} as
\begin{align}
S_{\rm free} & = \frac12 \int dt \, dy \, dt' \, dy' \, \phi(y, t) K(y,y'; t,t') \phi(y', t') \equiv \frac12 \phi \cdot K \cdot \phi \\
S_{\rm int} & = \frac12 \int dt \, dy \, dt' \, dy' \, \phi(y, t) V(y,y'; t,t') \phi(y', t') \equiv \frac12 \phi \cdot V \cdot \phi \; ,
\end{align}
where the dot stands for infinite-dimensional matrix-multiplication, which thus includes the spacetime integrals, and  $K$ (for `kinetic') and $V$ (for `vertex') are infinite-dimensional, hermitian matrices  defined as
\begin{align}
K = &\left(\partial_t \partial_t' - \partial_y \partial_y'\right)\delta(y' - y)\delta(t' - t),\\
V = &\left(\partial_t \partial_y' + \partial_t' \partial_y\right)\left(H(t)y\,\delta(y' - y)\delta(t' - t)\right)\\
& + \partial_y \partial_y'\left(H^2(t) y^2 \, \delta(y' - y)\delta(t' - t)\right).
\end{align}
We are displaying these expressions for completeness, but we will not need them in what follows.

The two-point function \autoref{2pt} must be a Green's function for the operator $K+V$. That is, it must obey the equation
\beq \label{Deom}
(K + V)\cdot D = i \, \mathbbm{1} \; ,
\eeq
where the identity matrix in this context is $\mathbbm{1} = \delta(y-y') \delta(t-t')$.

Let's now  instead see what perturbation theory yields. Treating the interactions in $V$ as perturbations, we would write the two-point function as an infinite sum of Feynman diagrams,
%
%
%
%
%
    \begin{equation}
    D_{\rm pert}~ = ~
 \feynmandiagram [baseline=-\the\dimexpr\fontdimen22\textfont2\relax,small, layered layout, horizontal'= a to b] { a --  b };
~+~
\feynmandiagram[baseline=-\the\dimexpr\fontdimen22\textfont2\relax,small, layered layout, horizontal = a to b]
{a  --  b [empty dot, minimum size=2mm]  --  c};
~+~
\feynmandiagram[baseline=-\the\dimexpr\fontdimen22\textfont2\relax,small, layered layout, horizontal = a to b]
{a  --  b [empty dot, minimum size=2mm]  --  c [empty dot, minimum size=2mm] -- d};
~+~\ldots \, ,
\label{PropSeries}
\end{equation}
where with the line we denote a free propagator, and now the empty dot stands  for the sum of all  interaction vertices contained in $V$: since we are now looking at the whole perturbative series, there is really no need to differentiate between vertices of different orders in $H$. Notice that, for now, we are calling $D_{\rm pert}$ the two-point function that we compute in perturbation theory, leaving open the possibility that it might differ from the full, non-perturbative one \autoref{2pt} by non-perturbative terms.

In formulae, as usual, we have a geometric-looking series:
\begin{align}
D_{\rm pert} & = D_0 + D_0 \cdot i V \cdot D_0 + D_0 \cdot i V \cdot D_0 \cdot i V \cdot D_0 + \dots  \label{series} \\
& = D_0 \cdot \big(\mathbbm{1} + i V \cdot D_0 + i V \cdot D_0 \cdot i V \cdot D_0 + \dots \big ) \; , \nonumber
\end{align}
where $D_0$ is the free propagator, which is a Green's function for the kinetic operator $K$,
\beq \label{D0}
K \cdot D_0 = i \, \mathbbm{1} \;. 
\eeq
However, in our case we are not actually able to sum this series. In the usual case, thanks to momentum conservation, in momentum space the series is just a sum of {\em numbers}, or at most finite-dimensional matrices (if one has more fields or fields with spin.) Here, instead, even in momentum space we have non-trivial differential operators and convolutions, and so taking inverses is not immediate. In other words, we could formally write
\beq
D_{\rm pert} = D_0 \cdot  \big( \, \mathbbm{1} - i V \cdot D_0  \, \big)^{-1} \; , 
\eeq
but that apparently harmless $(\dots)^{-1}$ actually stands for the inverse of an operator in an infinite-dimensional functional space, which we are not able to write down explicitly.

Regardless, the series \autoref{series} must obey the resolvent identity
\beq
D_{\rm pert} = D_0 +  D_0 \cdot i V \cdot D_{\rm pert} \; ,
\eeq
or, rearranging terms,
\beq
\big(\mathbbm{1} - D_0 \cdot i V \big)\cdot D_{\rm pert} = D_0 \; .
\eeq
Applying $K$ from the left to both sides, and using \autoref{D0}, we recover the equation of motion that the full, non-perturbative two-point function must obey, eq.~\autoref{Deom}, therefore $D_{\rm pert} = D$.

In conclusion, at least as far as the two-point function is concerned, our perturbative expansion correctly reproduces the full result, with no room for additional non-perturbative effects. Moreover, we restricted to $1+1$ dimensions and to the massless, minimally coupled case for notational simplicity, but clearly, the manipulations above are so general that they apply in any spacetime dimensions and for any quadratic Lagrangian. 
Then, even in the more general case, 
what one finds in perturbation theory matches  exactly the non-perturbative two-point function.
\section{From the two-point function to the transition amplitude}\label{AmplMatch}

We can go one step further and apply the LSZ reduction formula to our two-point function, to see if it correctly reproduces the one-to-one transition amplitude. For simplicity, we again restrict to the 1+1 dimensional, massless, minimally/conformally coupled case. So, the formula to reproduce is \autoref{1+1massless}, where  $\kappa$ is defined in \autoref{parametrization}.

In 1+1 dimensions, the LSZ reduction formula for a one-to-one process reads
\begin{align}
\label{LSZ}
\langle q | S | p \rangle = (-i)^2(\w_p^2 - p ^2)& (\w_q^2 - q^2) \times \\ \int d y \, d y' & \, d t \, d t' \, D(y, y'; t, t')e^{i p y}e^{-i q y'} e^{-i \w_p t}e^{i \w_q t'} \, , \nonumber
\end{align}
where the on-shell limit, 
\beq \label{on shell}
\w_p \to p \; , \qquad \w_q \to q \; ,
\eeq
is understood.

As emphasized in the previous section, in the case at hand the adiabatic expression for the field's two-point function, eq.~\autoref{PropFull}, is exact. We can then perform the integrals in $y$, $y'$, and $k$ right away. Using the parametrization in \autoref{parametrization} for $p$ and $q$ as well as a time-dependent generalization thereof for the scale factors,
\beq
A(t,t') \equiv \sqrt{a(t) a(t')} \; , \qquad N(t,t') \equiv \log \lb a(t')/a(t)\rb = \int_t^{t'} \! ds  \, H(s)\; ,
\eeq
we are left with
\begin{align}
\langle q | S | p \rangle = &  -(\w_p^2 - p ^2)(\w_q^2 - q^2) \times \\
& \int \frac{ d t \, d t' \, d \omega}{2\pi} \, \frac{i A(t,t')}{Q(\omega^2- k^2 + i \e)} e^{-i \omega (\t' -\t)} e^{-i \w_p t} e^{i \w_q t'} \, (2\pi )\d \big( \k - N(t,t') \big) \, ,  \nonumber
\end{align}
where $k$ is a placeholder for
\beq
k = a(t) p = a(t') q = A(t,t') Q \; .
\eeq

For simplicity, we can focus on the $\k > 0 $ ($p > q$) case only, which, as we already know, is the only possibility for an expanding cosmology. Then, the delta function enforces $t' > t$, which allows us to
close the contour for the $\omega$ integral in the lower half-plane. This yields
\begin{align}
\label{eq:LSZalmostthere}
\langle q | S | p \rangle = &  - \frac{(\w_p^2 - p ^2)(\w_q^2 - q^2)}{2 p q} \times \\
& \int d t \, d t' \,  e^{-i \w_p t} e^{i (k-i\e) \tau} e^{i \w_q t'} e^{-i (k-i\e) \tau'} \, (2\pi )\d \big( \k - N(t,t') \big) \,   
\theta(t' - t) \; ,\nonumber
\end{align}
where we have kept the $i \e$'s at the exponent because they are important for what follows.

Now we remember that we have to take the on-shell limit \autoref{on shell}. The prefactor of the integral vanishes in that limit
\beq
 \frac{(\w_p^2 - p ^2)(\w_q^2 - q^2)}{2 p q} \to 2 (\w_p - p  )(\w_q - q) \to 0 \; ,
\eeq
while the {\em integrand} is regular. Thus, any nonzero contributions can  only come from the regions of integration at infinity, which together with the condition $t'>t$ ensures that only the $t \to -\infty$, $t' \to + \infty$ regions can deliver the required  divergence.

In those limits, we simply have
\beq \label{limits}
a(t) \to a_1 \; , \qquad \tau \to c_1 + \frac{t}{a_1} \; , \qquad a(t') \to a_2 \; ,\qquad \tau' \to c_2 + \frac{t'}{a_2} \;, 
\eeq
where $c_1 $ and $c_2$ are integration constants that we will determine shortly. Performing the phase integrals in those limits and focusing on energies close to on-shell, we acquire the requisite energy poles:
\begin{align}
\left.  \int_{-\inf}^{t_0} dt\, e^{i (p - \w_p - i \e) t}\; \right|_{\w_p \to p} & = \f{i}{\w_p - p + i \epsilon } + \mbox{non-singular,}\\
\left. \int_{t'_0}^{\infty} dt'\, e^{-i (q - \w_q - i \e) t'} \; \right|_{\w_q \to q} & = \f{i}{\w_q - q + i \epsilon } + \mbox{non-singular,}
\end{align}
where $t_0$ and $t_0'$ are arbitrary finite times. We thus get
\begin{align}
\langle q | S | p \rangle =\, &  2  \, e^{ - i Q A \, (c_2- c_1)}(2\pi )\d \big( \k - N \big) \; ,\nonumber
\end{align}
where $N = \tilde H(0)$ is now the total number of $e$-folds, and $A$ is the common scale factor in \autoref{parametrization}. Apart from the phase, the transition amplitude is clearly the same as in \autoref{1+1massless}. 

We now prove that the phases also match: from \autoref{limits}, we have
\begin{align}
Q A\,  (c_2- c_1 ) & = \lim_{t' \to \infty} Q A \Big( \tau(t') - \frac{t'}{a(t')} \Big) - \lim_{t \to -\infty} Q A \Big( \tau(t) - \frac{t}{a(t)} \Big)  \\
& = QA \int_{-\infty} ^\infty \! dt  \, \Big(\frac{1}{a(t)} -  \frac{d}{dt} \frac{t}{a(t)} \Big) \\
& = -QA \int_{-\infty} ^\infty \! dt  \, t \, \frac{d}{dt} \frac{1}{a(t)} \\
& = - \int_{-\infty} ^\infty \! dt  \, t \, \dot \omega(t) \; , 
\end{align}
where in the last line we recalled $\omega(t) = Q A/a(t)$. The resulting expression reproduces the correct phase, expressed as in \autoref{phase t omegadot}.

\section{Concluding remarks}
We have laid the foundations for a perturbative approach to cosmological effects in quantum field theory. Notice the order of the terms: we take QFT in flat space as a starting point, and we treat the cosmological expansion as a small perturbation, rather than the other way around (small quantum effects in cosmology). This is a sensible viewpoint for distances much shorter than the Hubble length.

Much remains to be understood in terms of the systematics of this perturbative expansion, in particular, how to make it more directly in terms of $H/k$, with $k$ being the physical wavenumber, rather than in terms of $\tilde H$, which, if evaluated at zero frequency, is the total number of $e$-folds---hardly something we want to be expanding in in realistic situations.

Still, our sample computations already confirm something that was expected on physical grounds: for field modes that are much shorter than the Hubble radius, cosmology {\em is} a weak, adiabatic external gravitational field. The familiar phenomena of cosmological redshift  and particle creation can be reliably computed in perturbation theory, diagramatically, treating this weak gravitational field as an external source coupled to our quantum field. There is no deep geometric phenomenon, no mysterious ``stretching'' of space going on, no curved-space subtleties about the concept of particle. For the two-point function of our quantum field, we were also able to show that resumming our perturbative expansion yields the full result, with no room for additional non-perturbative phenomena. 

To be clear, we restricted ourselves to the simplest possible case: a free scalar field, living in an FRW cosmology with no gravitational backreaction. Nevertheless, it seems to us that the technical complications that giving up one or more of these assumptions would introduce, would not affect the main point of our analysis. In particular, we introduced a system of coordinates where the cosmological expansion   corresponds manifestly to a weak, adiabatic external source. In principle, in the same coordinates, we could add interactions and/or other fields to our QFT, and introduce gravitational backreaction as well. We would then have a more complicated perturbative expansion, but the cosmological expansion would still correspond to weak, adiabatic source, whose effects we could compute in perturbation theory.

In light of all this, it is difficult to find room for the trans-Planckian problem. As a crude toy model for trans-Planckian physics, consider for example a heavy scalar $X$ weakly coupled to a light scalar $\phi$ in a generic way. Let us use  our physical coordinates $(t, \vb{y})$, and let us restrict to cosmologies of the type discussed in sect.~\ref{s:PerturbativeCalcs}, for which we have a well defined notion of vacuum state in the far past and in the far future. If the system starts in the far-past vacuum state,
there are two types of ``trans-Planckian'' effects we might  worry about. 

The first is the production of quanta of the heavy $X$ field. As we have seen in sect.~\ref{s:PerturbativeCalcs}, this can be reliably computed in perturbation theory, and, to leading order, the production probability per unit phase-space volume happens to be at most of order  $|\, \tilde H(2 M_X) \, |^2$, where $M_X$ is the mass of our $X$ particles.   If $H(t)$ and its time derivatives are very small in units of $M_X$, this probability is extremely small. For instance, for a Gaussian $H(t)$ with height $H_0 \ll M_X$ and width $H_0^{-1}$,  such a probability is exponentially small, of order $e^{- 4 (M_X/H_0)^2}$. One can thus safely assume that, to extremely good accuracy, $X$ remains in its vacuum state. This is the quantum mechanical adiabatic theorem at work.

The second effect has to do with how the interactions of $\phi$ with $X$ affect the correlation functions of $\phi$ itself as well as the production of $\phi$ quanta. But, since there are no quanta of $X$ around in the far past or in the far future, this is {\em completetely} captured by the low-energy effective field theory in which $X$ has been integrated out: such a theory will involve self-interactions of $\phi$, organized in a derivative expansion in which higher and higher orders are suppressed by higher and higher powers of $M_X$. 
We all believe that trans-Planckian physics is in principle encoded in higher-dimensional operators in our low-energy effective theories, and so this second effect is nothing new (nor are we the first to realize this---see e.g.~\cite{KKLS}). 

We hope that our explicit perturbative setup and calculations will provide a starting point to address these questions more concretely and systematically.

\acknowledgments
We thank Paolo Creminelli, Lam Hui, Alessandro Podo, and Andrew Tolley for useful discussions.
We are especially thankful to Rafael Krichevsky and Federico Piazza for collaboration in the early stages of this project. Our work is partially supported by the US DOE (award number DE-SC011941) and by the Simons Foundation (award number 658906).

\newpage

\appendix

\section{Computations using QFT in curved spacetime}
\label{app:TextbookResults}

In this section, we review the usual approach to doing QFT in a cosmological background. Throughout this appendix, we will be using the standard FRW form of the metric with $\left(t, \vb{x}\right)$ co-moving coordinates, whose line element is given in \autoref{eq:FRW}. The methods used in this section are studied extensively in refs. \cite{ParkerToms, BirrellDavies}. We begin in \autoref{app:ParticleProduction} by deriving particle production for the scale factor referenced in \autoref{s:ParticleProduction}. Then, in \autoref{app:GravitationalRedshift}, we derive the transition amplitude that is matched in \autoref{s:GravitationalRedshift}.

\subsection{Particle Production}
\label{app:ParticleProduction}
A massless minimally coupled scalar field in a curved background has as its action
\begin{equation}
    S = -\frac{1}{2}\int dt d^n x \, g^{\mu\nu}\partial_\mu \phi \partial_\nu \phi \sqrt{-g} .
\end{equation}
Using FRW coordinates, the equation of motion is
\begin{equation}
    \f{d^2 \tilde{\phi}_{\vb{k}}}{d \tau^2} + a^{2\left(n - 1\right)}\left(\tau\right) k^2 \tilde{\phi}_{\vb{k}} = 0,
\end{equation}
where $\tilde{\phi}_{\vb{k}}\left(\tau\right)$ are the spatial Fourier modes of $\phi\left(x, \tau\right)$, $\vb{k}$ is the {\em comoving} momentum,  and $d {\tau} = d{t}/a^n$. We take the scale factor to be the $n$-dimensional generalization of what was studied in ref. \cite{BirrellDavies},
\begin{equation}
    a^{2\left(n - 1\right)}\left(\tau\right) = a_1^{2\left(n -1\right)} + \frac{e^{\tau/s}}{\left(e^{\tau/s} + 1\right)^2}\left[\left(a_2^{2\left(n - 1\right)} - a_1^{2\left(n - 1\right)}\right)\left(e^{\tau/s} + 1\right) + b\right].
\end{equation}
Note that the scale factor has the property that it approaches $a_1$ in the infinite past and $a_2$ in the infinite future, as desired. We can substitute $u = e^{\tau/s}$, and so the differential equation for $\tilde{\phi}_{\vb{k}}\left(u\right)$ is,
\begin{equation}
    \f{d^2\tilde{\phi}_{\vb{k}}}{d u^2} + \frac{1}{u}\f{d  \tilde{\phi}_{\vb{k}}}{d u} + k^2 s^2\left(\frac{a_1^{2\left(n - 1\right)}}{u^2} + \frac{b + \left(a_2^{2\left(n - 1\right)} - a_1^{2\left(n - 1\right)}\right)\left(u + 1\right)}{u\left(u + 1\right)^2}\right)\tilde{\phi}_{\vb{k}} = 0.
\end{equation}
This has 3 regular singularities, meaning that it will eventually be solved by hypergeometric functions. The general strategy for finding solutions at this point is to apply a M\"{o}bius transformation to shift the singularities over to $(0, 1, \infty)$ and then to apply an indicial transformation so as to write the equation into its asymmetrically reduced form. In that form, we can identify what special function we are dealing with and its parameters. In this case, the M\"{o}bius transformation is linear, $z = u + 1$, making the substitution simple,
\begin{equation}
    \f{d^2 \tilde{\phi}_{\vb{k}}}{d z^2} - \frac{1}{1 - z}\f{d \tilde{\phi}_{\vb{k}}}{d z} + k^2 s^2\left(\frac{a_1^{2\left(n - 1\right)}}{\left(1 - z\right)^2} - \frac{b + \left(a_2^{2\left(n - 1\right)} - a_1^{2\left(n - 1\right)}\right)z}{z^2\left(1 - z\right)}\right)\tilde{\phi}_{\vb{k}} = 0.
\end{equation}
The indicial transformation comes with a sign ambiguity. We choose the sign such that the solution in the infinite past has the expected positive frequency mode form,
\begin{equation}
    \tilde{\phi}_{\vb{k}}\left(z\right) = z^{r_0} \left(1 - z\right)^{-r_1}h\left(z\right),
\end{equation}
where $r_0$, $r_1$, and $r_2$ (for future convenience) are defined as
\begin{equation}
    r_0 = \frac{1}{2}\left(1 - \sqrt{1 + 4k^2 s^2 b}\right)\; , \qquad r_1 = iksa_1^{n - 1} \; , \qquad r_2 = iksa_2^{n - 1} \; ,
\end{equation}
where $k = \abs{\vb{k}}$. Effecting this transformation ensures that $h\left(z\right)$ satisfies the standard differential equation for a Gauss hypergeometric function, $F$,
\begin{equation}
    \f{d^2 h}{d z^2} + \left(\frac{2r_0}{z} - \frac{1 - 2r_1}{1 - z}\right)\f{d h}{d z} - \frac{\left(r_0 - r_1\right)^2 - r_2^2}{z\left(1 - z\right)}\, h = 0.
\end{equation}
From here, we can identify the hypergeometric function parameters as $a = r_0 - r_1 - r_2$, $b = r_0 - r_1 + r_2$, and $c = 2r_0$. The singularity at $1$ corresponds to the infinite past, so the positive-frequency normalized mode function for the infinite past is,
\begin{equation}
    \tilde{\phi}_{\vb{k}}^{\text{in}} = \frac{1}{\sqrt{2k a_1^{n - 2}}}z^{r_0}\left(z - 1\right)^{-r_1} F\left(r_0 - r_1 - r_2, r_0 - r_1 + r_2; 1 - 2r_1; 1 - z\right)
\end{equation}
The singularity at $\infty$ corresponds to the infinite future, so the positive-frequency normalized mode function for the infinite future is,
\begin{equation}
    \tilde{\phi}_{\vb{k}}^{\text{out}} = \frac{1}{\sqrt{2ka_2^{n-2}}}z^{r_1 - r_2} \left(z - 1\right)^{-r_1} F\left(r_0 - r_1 + r_2, 1 - r_0 - r_1 + r_2; 1 + 2 r_2; \frac{1}{z}\right)
\end{equation}
Upon using the following connection formula \citep[Eq. 15.10.27]{Hypergeometric},
\begin{equation}
\begin{aligned}
    F\left(a, b, a + b - c + 1; 1 - z\right) = \\
    &\hspace{-2cm}\frac{\Gamma\left(a + b - c + 1\right)\Gamma\left(b - a\right)}{\Gamma\left(b\right)\Gamma\left(b - c + 1\right)}z^{-a} F\left(a, a - c + 1, a - b + 1, \frac{1}{z}\right) + \\
    &\hspace{-2cm}\frac{\Gamma\left(a + b - c + 1\right)\Gamma\left(a - b\right)}{\Gamma\left(a\right)\Gamma\left(a - c + 1\right)} z^{-b} F\left(b, b - c + 1, b - a + 1, \frac{1}{z}\right).
\end{aligned}
\end{equation}
along with $\tilde{\phi}_{\vb{k}}^{\text{in}} = \alpha_k \,  \tilde{\phi}_{\vb{k}}^{\text{out}} + \beta_k \, \tilde{\phi}_{-\vb{k}}^{\text{out}\: \ast}$, we find that the $\beta$ Bogolyubov coefficients are
\begin{equation}
\label{eq:BogolyubovFinal}
    \beta_k = \left(\frac{a_2}{a_1}\right)^{\frac{n}{2} - 1}\frac{\Gamma\left(1 - 2r_1\right)\Gamma\left(2r_2\right)}{\Gamma\left(r_0 - r_1 + r_2\right)\Gamma\left(1 - r_0 - r_1 + r_2\right)} \; .
\end{equation}
There are two simplifications that are used in the main text. First, when $b = 0$, the scale factor becomes that of \autoref{eq:ParticleEx1}. In that case, we obtain
\begin{equation}
    \beta_k = \left(\frac{a_2}{a_1}\right)^{\frac{n}{2} - 1}\frac{\Gamma\left(1 - 2iksa_1^{n - 1}\right)\Gamma\left(iksa_2^{n-1}\right)}{\Gamma\left(iks\left(a_2^{n - 1} - a_1^{n-1}\right)\right)\Gamma\left(1 + iks\left(a_2^{n - 1} - a_1^{n - 1}\right)\right)} \; .
\end{equation}
At this point, we can let $a_2 = a_1 + \Delta a$ and expand to first order in $\Delta a/a_1$. It is helpful to simplify the result by using the Euler reflection formula,
\begin{equation}
    \Gamma\left(z\right)\Gamma\left(1 - z\right) = \frac{\pi}{\sin\left(\pi z\right)}.
\end{equation}
The result expanded to first order in $\Delta a/a_1$ is
\begin{equation}
\label{eq:exactex1}
    \beta_k \simeq \frac{1}{2}\left[\left(n - 1\right)\frac{2\pi s a_1^{n - 1} k}{\sinh\left(2\pi s a_1^{n - 1} k\right)} \frac{\Delta a}{a_1}\right].
\end{equation}
Note that in the main text, we write it in terms of the physical momentum $p = k/a_1$. Second, when $a_2 = a_1$, the scale factor becomes that of \autoref{eq:ParticleEx2}. The Bogolyubov coefficient is then
\begin{equation}
    \beta_k = \frac{\Gamma\left(1 - 2iksa_1^{n - 1}\right) \Gamma\left(2iksa_1^{n - 1}\right)}{\Gamma\left(r_0\right)\Gamma\left(1 - r_0\right)} \; .
\end{equation}
Once more we can use the Euler reflection formula, but this time we expand to first order in $b$. In doing so, we obtain
\begin{equation}
\label{eq:exactex2}
    \beta_k \simeq \frac{1}{2}\left[\frac{2\pi s^2 k^2}{\sinh\left(2\pi s a_1^{n - 1} k\right)} b\right].
\end{equation}

\subsection{Gravitational Redshift}
\label{app:GravitationalRedshift}
We now want to derive the $S$-matrix element describing the gravitational redshift of a single-particle state for our scalar field. We will do so in the adiabatic approximation, where we can  solve the theory analytically for generic scale factors. Recall that the adiabatic limit for us corresponds to modes that never leave the horizon, so that the cosmological expansion is, for them, adiabatically slow. 

The set-up is simple: at all times, the field operator can be expanded as
\beq \label{field expansion}
\phi(\vb{x}, t) = \int \frac{d^n k}{(2 \pi)^n} \big[ b_{\vb{k}} \, u_{\vb{k}}(\vb{x},t) + \rm{h.c.} \big] \; ,
\eeq
where $\vb{k}$  stands for {\em comoving} momentum, and the $u_{\vb{k}}$'s and their complex conjugate form a complete set of solutions of the field equations, orthonormalized with respect to the scalar product
\beq
\langle f,g \rangle = i \int \! d^n x \, a^n(t) f^*(\vb{x},t) \,  \overleftrightarrow{\partial_t} g(\vb{x},t) \; . 
\eeq
In cases when there is no ambiguity about positive vs.~negative frequency modes, one takes the $u_{\vb{k}}$'s as the positive frequency ones, and then the $b_{\vb{k}}$'s have the usual interpretation as annihilation operators \cite{BirrellDavies}.

In the adiabatic limit, we can use the WKB approximation and write the positive frequency solutions as usual as $\exp (- i \int \omega(t) dt)$, where $\omega(t)$ is the instantaneous frequency associated with comoving momentum $\vb{k}$. With the correct normalization for the scalar product above, and choosing an arbitrary initial phase, we have
\beq
u_{\vb{k}}(y,t) = \frac{1}{\sqrt{2 \, \omega(t) \, a^n(t)}} \, e^{- i \int_0^t dt' \, \omega(t')} \, e^{  i \vb{k} \cdot \vb{x}} \; ,
\eeq 

On the other hand, from the viewpoint of the flat spacetime one ends up with in the infinite past and in the infinite future, the field expansion must be the usual Minkowski space one, with different annihilation operators $b^{\rm in/out}_{\vb{p}}$:
\beq
\phi(\vb{x}=\vb{y}/a(t), t) \big|_{t\to{\mp \infty}} = \int \frac{d^n p}{(2 \pi)^n} \, \frac{1}{\sqrt{2 \, \omega_{p}}} \big[ \,  b^{\rm in/out}_{\vb{p}} \, e^{- i \omega_{{p}} t + i \vb{p}\cdot \vb{y}} + \rm{h.c.} \big] \; ,
\eeq
where $\vb{p}$ stands for {\em physical} momentum and $\vb{y}$ for physical coordinates, and $\omega_{{p}} = \sqrt{\vb{p}^2+m^2}$ is the usual relativistic energy. Equating this with the early and late time limits of \eqref{field expansion}, we find the relationships between the adiabatic annihilation operators and the asymptotic ones:
\begin{align}
b_{\vb k} 
 & = a_1^{n/2} e^{-i \varphi_{\rm in}}  \, b^{\rm in}_{\vb{p= \vb{k}/a_1}}  \\
& = a_2^{n/2} e^{i \varphi_{\rm out}} \, b^{\rm out}_{\vb{q= \vb{k}/a_2}}
\end{align}
where the phases are defined as
\beq
\varphi_{\rm in}= \int^0_{-\infty} (\omega(t)- \omega_p) dt \; , \qquad  \varphi_{\rm out} = \int_0^\infty (\omega(t)- \omega_q) dt \; .
\eeq
Notice that, at this order in the WKB approximation, there is no mixing between positive and negative energy modes, and so the $\beta$ Bogolyubov coefficients vanish.

Now, our asymptotic states, with the usual relativistic normalization, are simply
\beq
| \vb{p}\rangle_{\rm in}= \sqrt{2 \omega_p} \, b^{\dagger \,{\rm in}}_{\vb{p}} |0 \rangle \; , \qquad | \vb{q}\rangle_{\rm out}= \sqrt{2 \omega_q} \, b^{\dagger \,{\rm out}}_{\vb{q}} |0 \rangle \; .
\eeq
Putting everything together, we find the desired $S$-matrix element:
\begin{align}
{}_{\rm out} \langle \vb{q} | \vb{p} \rangle_{\rm in} & =e^{-i(\varphi_{\rm in}+\varphi_{\rm out}) } \sqrt{2 \omega_p}\sqrt{2 \omega_q}(a_1 a_2)^{n/2} \langle 0 | b_{{a_2 \vb{q}}} \, b^\dagger_{a_1 \vb{p}}|0 \rangle \; ,
\end{align}
which, using $[b_{\vb{k}}, b^\dagger_{\vb{k}'}] = (2\pi)^n \delta^n(\vb{k}-\vb{k}')$, precisely yields \autoref{full redshift}.

\section{Matching the transition amplitude to \texorpdfstring{2$^{\text{nd}}$}{2nd} order}
\label{App1to1PertCom}

Our goal in this section is to demonstrate the second-order matching of the two expressions for the $1 \to 1$ transition amplitude in $(1+1)D$ that we obtained earlier. We restrict to the massless, minimally coupled case for simplicity. The first expression, \autoref{full redshift} or, equivalently, \autoref{final amplitude 1+1}, was derived non-perturbatively, whereas the second was obtained via the Feynman rules---see \autoref{general n-th order} with $j=2$.

We begin with the expansion of the full $1 \to 1$ scattering amplitude given by \autoref{full redshift}
\beq
\braket{q|S|p} = 2 \sqrt{pq} e^{- i \vp} \sqrt{a_1 a_2}\,(2 \pi)\, \d (a_2 q - a_1 p) \, .
\label{ampPhysApp}
\eeq
The expansion in $\tih$ would involve the phase
\beq
\vp = - \int \limits d t \, t \, \dot\w(t) \; , 
\label{phasePhysApp}
\eeq
which we deal with first. The linear-in-$\tih$ term in the expansion of the phase was obtained earlier in \autoref{first order phase}, so we concentrate on $O(\tih^2)$:
\beq
\begin{aligned}
\vp^{(2)} & = - p \int d t \, d t' \, t H(t)  \theta (t-t') H(t') \\
& = - p \int  \, \f{dt \, d\w}{2  \pi} \, t H(t) \tilde \theta(\w) \tilde H(\w) e^{-i \w t} \\
& = - i p \int \frac{d \w}{2 \pi} \f{\tih(\w) \tih'(-\w)}{\w + i \e} \, .
\label{phase2nd1}
\end{aligned}
\eeq
Integrating by parts, we obtain
\beq
\vp^{(2)} = \frac{i p}{2} \tih(0) \tih' (0) + \f{p}{2} \int  \f{d \w}{2 \pi}\,  \f{|\tih(\w) |^2}{(\w + i \e)^2} \, .
\eeq
With this result, we can now find the $O(\tih^2)$ piece of the scattering amplitude \autoref{ampPhysApp} by expanding also the scale factors and the delta function to the same order
\begin{multline}
\braket{q|S|p}^{(2)} = 2 \pi p \, \d(q-p) \left\{  \tilde H^2(0) + 3 p \tilde H(0) \tilde H'(0) +  p^2 \tilde H'^2(0) - i p \Delta^+ \right\} \\
-2 \pi p^2 \, \d' (q-p) \left \{ 3 \tilde H^2(0)  + 2 p  \tilde H(0) \tilde H'(0)\right \} 
+2 \pi p^3 \tilde H^2(0) \, \d'' (q-p) \, ,
\label{ampl2}
\end{multline}
where we defined
\beq
\Delta^{+} \equiv \int  \f{d \w}{2 \pi}\,  \f{|\tih(\w)|^2}{(\w + i \e)^2} \, .
\eeq

The expression \autoref{ampl2} is to be compared with the $j=2$ result \autoref{general n-th order} computed using the Feynman rules
\begin{multline}
\braket{q | S | p}^{(2)} = 2 p q   \int  \frac{ d\omega_1 d k_1}{(2 \pi)^2} \frac{ i k_1}{\omega_1 - k_1 + i \epsilon \, {\rm sign}(k_1)}  \tilde H(q - \omega_{1})  \tilde H(\omega_1 - p) \\ \times (2 \pi)^2 \d'(q-k_{1}) \d'(k_{1} - p )  \; . 
\end{multline}
In the above, we perform the integral over $k_1$, and introduce a new integration variable $\w = \w_1-p$, which gives
\begin{multline}
\braket{q | S | p}^{(2)} = - 2 i p q \int d \w \f{p+\w}{(\w + i \e)^2} \tilde H(\w) \tilde H(q-p-\w) \, \d'(q-p) \\
+ 2 i p^2 q \int \f{d \w}{\w + i \e} \tilde H(\w) \tilde H(q-p-\w) \, \d''(q-p) \; ,
\end{multline}
where we removed the ambiguity in the direction of the pole shift by sticking with the convention $p>0$, ${\rm sign}\,(p)=1$. The rest of the check is just the matter of using the usual delta-function calculus, i.e. the identities \autoref{delta1} and \autoref{delta2}, to distribute the derivatives acting on $\d(q-p)$. This produces quite a long expression, which can afterwards be made more compact by using the identities
\beq
\bal
\int \frac{d\omega}{\omega + i \epsilon}  \tilde H(\omega) \tilde H(-\omega)& = - i \pi \tilde H^2(0)\, ,\\
\int \frac{d\omega}{\omega + i \e}  \tilde H(\omega) \tilde H'(-\omega) & = - i \pi \tilde H(0) \tilde H'(0) -  \pi \De^+ \, ,\\
\int \frac{d \omega}{\omega + i \e} \tilde H(\omega) \tilde H''(-\omega) & = - i \pi \tilde H'^2(0) - \int \frac{d \omega}{(\omega + i \e)^2} \tilde H(\omega) \tilde H'(- \omega)\, .
\eal
\eeq
 After some effort, one arrives at the exact same answer \autoref{ampl2} that we derived from the non-perturbative result. 

\section{The auxiliary field}
\label{app:auxiliary}

In \autoref{s:HigherOrders}, we showed how to extend our sub-horizon EFT beyond first order in perturbation theory. We emphasized that it is combinatorically difficult to find the $n^\text{th}$ order in $\tilde{H}$ term in the expansion of the $S$-matrix since both $O(\tilde{H})$ and $O(\tilde{H}^2)$ vertices are present in the theory. In this appendix, we demonstrate how one can rewrite the theory 
using an auxiliary field so that the interaction vertices are all manifestly first order in $\tilde{H}$, and the organization of the perturbative series is more natural.
For notational simplicity we restrict to the minimally coupled, massless, 1+1-dimensional case, but everything we say here can be straightforwardly extended to the more general case.


Consider the action \autoref{eq:Sfree}, \autoref{eq:Sint}. Introducing an auxiliary field $A$, it can be rewritten as 
\begin{equation}
\label{phiAact}
    S[\phi, A] = -\frac{1}{2}\int dt dy \, \left(\eta^{\mu\nu} \partial_\mu \phi \partial_\nu \phi + A^2 - 2Hy\partial_y \phi \left(\partial_0 \phi + A\right)\right).
\end{equation}
The equation of motion for the auxiliary field $A$ is simply $A = Hy\partial_y \phi$. If we insert this back into \autoref{phiAact}, we recover the original action. These manipulations are allowed at the quantum level as well, since the action is at most quadratic in $A$.

It is manifest that the theory now only contains terms that are at most linear in $\tilde{H}$. The price to pay is that we are dealing with two fields instead of one. To perform perturbative computations using this theory, it is helpful to switch to a matrix notation:
\begin{equation}
    \begin{aligned}
    \psi^a &= \begin{pmatrix} \phi \\ A \end{pmatrix}\\
    K_{ab} &= \begin{pmatrix} -\eta^{\mu\nu} \partial_\mu \partial_\nu & ~~ 0 \\ 0 & ~~ 1 \end{pmatrix} \hspace{0.15cm}\\
    V_{ab} &= \begin{pmatrix} H\partial_0 + 2Hy\partial_0 \partial_y + \dot{H}y\partial_y & \qquad H\left(1 + y\partial_y\right) \\ -Hy\partial_y & \qquad 0 \end{pmatrix} .
    \end{aligned}
\end{equation}
Notice that the $K$ and $V$ operators defined above are hermitian---hence the need for the $\dot H$ term in $V_{11}$. 

The new action then can be written in a simpler form,
\begin{equation}
    \begin{aligned}
    S[\psi] = -\frac{1}{2}\int dt dy \left(\psi^a K_{ab} \psi^b + \psi^a V_{ab} \psi^b\right).
    \end{aligned}
\end{equation}
Additionally, written in this way, the equation of motion is
\begin{equation}\label{eq:EOMAuxiliary}
    \left(K_{ab} + V_{ab}\right)\psi^b = 0\, .
\end{equation}
Using the Feynman rules, the interaction vertex can be written in terms of the interaction matrix, $V_{ab}$,
\begin{equation}
\vcenter{\hbox{
\begin{tikzpicture}
\feynmandiagram[horizontal = a to c, layered layout]
{a -- b [crossed dot] --c};
\end{tikzpicture}}\vspace{-0.3cm}}~~ = -i \tilde{V}_{ab},
\end{equation}
Explicitly, the components of this Fourier transform are given as follows,
\begin{equation}
\begin{aligned}
\tilde{V}_{11} &= 2\pi i\tilde{H}\left(\w_q - \w_p\right)\left[-\left(\w_q - \w_p\right)\delta\left(q - p\right) + \left(\w_p + \w_q\right)\left(p + q\right)\delta'\left(q - p\right)\right],\\
\tilde{V}_{12} &= 2\pi \tilde{H}\left(\w_q - \w_p\right)\left(2\delta\left(q - p\right) - \left(q - p\right)\delta'\left(q - p\right)\right),\\
\tilde{V}_{21} &= -2\pi \tilde{H}\left(\w_q - \w_p\right) \left(q - p\right) \delta'\left(q - p\right),\\
\tilde{V}_{22} &= 0\, .
\end{aligned}
\end{equation}
As a sanity check, the $\phi \to \phi$ transition amplitude that we computed earlier to first order in \autoref{firstordershift1} is correctly reproduced in this formalism by 
\begin{equation}
\mel{q}{S}{p}^{(1)} = -i\tilde{V}_{11}\, ,
\end{equation}
upon using the identity for the derivative of the Dirac delta function \autoref{diracid}.

Notice that while the perturbation series nicely organizes every term by their powers in $\tilde{H}$, it does not provide a computationally simpler way to handle higher order terms. For example, at $3^{\text{rd}}$ order for a $\phi \to \phi$ process, we would have to consider the $\phi \to \phi \to \phi \to \phi$, $\phi \to A \to \phi \to \phi$, $\phi \to \phi \to A \to \phi$, and $\phi \to A \to A \to \phi$ diagrams. To our knowledge, there is no general way to handle this matrix multiplication in an easy way.


As a final point, in this language, the fully dressed propagator is a matrix $G^{ab}$ with the quantity that we are interested in being $G^{11}$ (corresponding to a $T$-ordered correlation function of two $\phi$-fields). Diagrammatically, the fully dressed matrix-valued propagator can be expressed as a series
\begin{equation}
\vcenter{\hbox{
\begin{tikzpicture}
\feynmandiagram[layered layout, horizontal = a to b]
{a -- [double] b};
\end{tikzpicture}}}
~= 
\vcenter{\hbox{
\begin{tikzpicture}
\feynmandiagram[layered layout, horizontal = a to b]
{a -- [plain] b};
\end{tikzpicture}}}
~+ 
\vcenter{\hbox{
\begin{tikzpicture}
\feynmandiagram[layered layout, horizontal = a to c]
{a -- [plain] b [crossed dot, minimum size=3.5mm] -- [plain] c};
\end{tikzpicture}}\vspace{-0.3cm}}
~+ 
\vcenter{\hbox{
\begin{tikzpicture}
\feynmandiagram[layered layout, horizontal = a to d]
{a -- [plain] b [crossed dot, minimum size=3.5mm] -- [plain] c [crossed dot, minimum size=3.5mm] -- d};
\end{tikzpicture}}\vspace{-0.3cm}} ~+~ \cdots.  
\end{equation}
The number of vertices here, once again, conveniently corresponds to powers in $\tih$. In formulae, we can write
\begin{equation}
\begin{aligned}
    G^{ab} &= \left(iK^{-1}\right)^{ab} + \left(iK^{-1}\right)^{ac}iV_{cd}\left(iK^{-1}\right)^{db} + \dots,\\
    &= \left(K^{-1}\right)^{ac}\sum_{n = 0}^{\infty}\left[\left(iV \cdot iK^{-1}\right)^n\right]^{cb}.
\end{aligned}
\end{equation}
In the above expression, $V$ is the same interaction matrix that appears in the action. The inverse kinetic energy operator (i.e., the free propagator) has two non-zero components,
\begin{equation}
\begin{aligned}
\left(iK^{-1}\right)^{11}\left(y, y'; t, t'\right) &= \int \frac{d \w d k}{\left(2\pi\right)^2} \frac{i}{\w^2 - k^2 + i\epsilon} e^{-i \w \left(t - t'\right)} e^{ik\left(y - y'\right)} ,\\
\left(iK^{-1}\right)^{22}\left(y, y'; t, t'\right) &= i \, \delta\left(t' - t\right)\delta\left(y' - y\right).
\end{aligned}
\end{equation}

At this point, the procedure for handling the two-point correlation function is precisely the same as before. The fact that $G^{ab}$ is expressed in terms of this geometric series means that it must obey the resolvent identity. This is equivalent to saying that $G^{ab}$ satisfies the Green's function equation, as shown in \autoref{2ptMatch}. In the special case of a $(1 + 1)$-dimensional massless scalar field that is both minimally and conformally coupled, it is possible to write down $G^{11}$ explicitly, \autoref{PropFull}. Lastly, the setup of the auxiliary field demands that if $G^{11}$ satisfies the Green's function equation, so does the rest of $G^{ab}$. As one can see, the essential argument here is exactly the same as that used in the main text. The advantage here is simply found in ordering the perturbation series in a clearer way.


\begin{thebibliography}{99}

\bibitem{MB}
J.~Martin and R.~H.~Brandenberger,
``The TransPlanckian problem of inflationary cosmology,''
Phys. Rev. D \textbf{63}, 123501 (2001)
[arXiv:hep-th/0005209 [hep-th]].

\bibitem{BBLV}
A.~Bedroya, R.~Brandenberger, M.~Loverde and C.~Vafa,
``Trans-Planckian Censorship and Inflationary Cosmology,''
Phys. Rev. D \textbf{101}, no.10, 103502 (2020)
[arXiv:1909.11106 [hep-th]].

\bibitem{WeinbergQM}
S.~Weinberg,
{\em Lectures on Quantum Mechanics},
Cambridge: Cambridge University Press (2005). 

\bibitem{BdAQ}
C.~P.~Burgess, S.~P.~de Alwis and F.~Quevedo,
``Cosmological Trans-Planckian Conjectures are not Effective,''
JCAP \textbf{05}, 037 (2021)
[arXiv:2011.03069 [hep-th]].

\bibitem{Polchinski}
J.~Polchinski,
``String theory and black hole complementarity,''
[arXiv:hep-th/9507094 [hep-th]].

\bibitem{HN}
L.~Hui and A.~Nicolis,
``Two-dimensional Lorentz invariance of spherically symmetric black holes,''
Phys. Rev. D \textbf{89}, no.6, 064009 (2014)
[arXiv:1402.6707 [hep-th]].

\bibitem{Parikh}
M.~K.~Parikh,
``New coordinates for de Sitter space and de Sitter radiation,''
Phys. Lett. B \textbf{546}, 189-195 (2002)
[arXiv:hep-th/0204107 [hep-th]].

\bibitem{ParkerToms}
L.~Parker, \& D.~Toms,
\textit{Quantum Field Theory in Curved Spacetime: Quantized Fields and Gravity}, 
Cambridge: Cambridge University Press (2009). 

\bibitem{BirrellDavies}
N.~Birrell,  \& P.~Davies,
\textit{Quantum Fields in Curved Space},
Cambridge: Cambridge University Press (1982). 

\bibitem{Weinberg1}
S.~Weinberg,
\textit{The Quantum Theory of Fields},
Cambridge: Cambridge University Press (1995). 

\bibitem{Gubser1}
S.~S.~Gubser,
``String production at the level of effective field theory,''
Phys. Rev. D \textbf{69}, 123507 (2004)
[arXiv:hep-th/0305099 [hep-th]].

\bibitem{Gubser2}
S.~S.~Gubser,
``String creation and cosmology,''
[arXiv:hep-th/0312321 [hep-th]].

\bibitem{Tolley2005}
A.~Tolley and D.~H.~Wesley, 
``String Pair Production in a Time-Dependent Gravitational Field", 
Phys. Rev. D, \textbf{72} 124009 (205) 
[arXiv:0509151 [hep-th]].


\bibitem{Hypergeometric}
\emph{NIST Digital Library of Mathematical Functions}, 
\url{http://dlmf.nist.gov/}, Release 1.1.5 of 2022-03-15, 
F.~W.~J.~Olver et al.~eds.



\bibitem{KKLS}
N.~Kaloper, M.~Kleban, A.~E.~Lawrence and S.~Shenker,
``Signatures of short distance physics in the cosmic microwave background,''
Phys. Rev. D \textbf{66}, 123510 (2002)
[arXiv:hep-th/0201158 [hep-th]].



\end{thebibliography}
\end{document}